# Attribute Oriented Induction with simple select SQL statement

**Spits Warnars**
*Department of Computing and Mathematics, Manchester Metropolitan University*
*John Dalton Building, Chester Street, Manchester M15GD, United Kingdom*
(*s.warnars@mmu.ac.uk*)

**Abstract:**
Searching learning or rules in relational database for data mining purposes with characteristic or classification/discriminant rule in attribute oriented induction technique can be quicker, easy, and simple with simple SQL statement. With just only one simple SQL statement, characteristic and classification rule can be created simultaneously. Collaboration SQL statement with any other application software will increase the ability for creating t-weight as measurement the typicality of each record in the characteristic rule and d-weight as measurement the discriminating behavior of the learned classification/discriminant rule, particularly for further generalization in characteristic rule. Handling concept hierarchy into tables based on concept tree will influence for the successful simple SQL statement and by knowing the right standard knowledge to transform each of concept tree in concept hierarchy into one table as transforming concept hierarchy into table, the simple SQL statement can be run properly.

1. INTRODUCTION

Attribute oriented induction approach is developed for learning different kinds of knowledge rules such as characteristic rules, discrimination or classification rules, quantitative rules, data evolution regularities [15], qualitative rules [9], association rules and cluster description rules [13]. Attribute oriented induction has concept hierarchy as an advantage where concept hierarchy as a background knowledge which can be provided by knowledge engineers or domain experts [8,10,11]. Concepts are ordered in a concept hierarchy by levels from specific or low level concepts into general or higher level and generalization is achieved by ascending to the next higher level concepts along the paths of concept hierarchy [11]. The attribute oriented induction method integrates a machine learning paradigm especially learning-from-examples techniques [3] with database operations, extracts generalized rules from an interesting set of data and discovers high level data regularities [13].

The attribute oriented induction method has been implemented in data mining system prototype called DBMINER [16,17] which previously called DBLearn [12,14] and been tested successfully against large relational database and datawarehouse for multidimensional purposes.

For the implementation attribute oriented can be implemented as the architecture design in figure 1 where characteristic rule and classification rule can be learned straight from transactional database (OLTP) or Data warehouse (OLAP) with the helping of concept hierarchy as knowledge generalization. Concept hierarchy can be created from OLTP database as a direct resource.

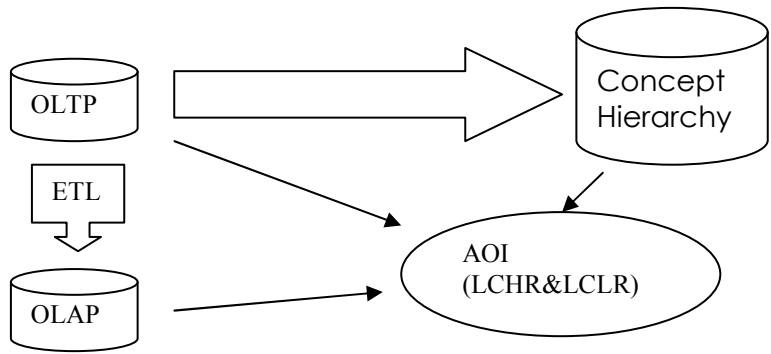

Fig. 1. AOI architecture

For making easy the implementation a concept hierarchy will just only based on non rule based concept hierarchy and just learning for characteristic rule and classification/discriminant rule.
1) Characteristic rule is an assertion which characterizes the concepts which satisfied by all of the data stored in database. Provide generalized concepts about a property which can help people recognize the common features of the data in a class. For example the symptom of the specific disease [2].
2) Classification/Discriminant rule is an assertion which discriminates the concepts of 1 class from other classes. Give a discriminant criterion which can be used to predict the class membership of of new data. For example to distinguish one disease from others, classification rule should summarize the symptoms that discriminate this disease from others [2].

From database we can learn 2 learning they are:
1) Positive learning as target class where the data are tuples in the database consistent with the learning concepts

2) Negative learning as contrasting class in which the data do not belong to the target class.

Thus positive learning/target class will be built when do characteristic rule and negative learning/contrasting class will be built when do classification rule where positive learning/target class as result of characteristic rule must be done previously. Figure 2 explain about this matter.

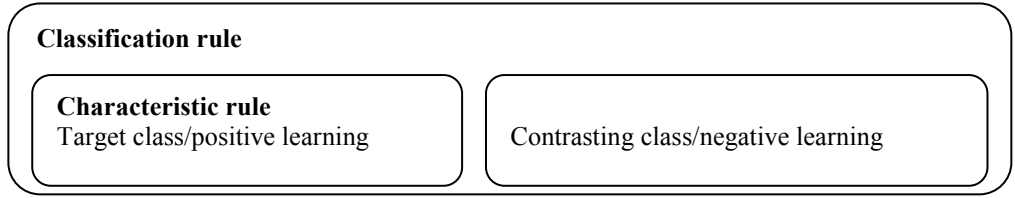

Fig. 2. Classification and characteristic rule

For doing the generalization there are 8 strategy steps must be done [8], where step 1 until 7 as for characteristic rule and step 1 until 8 as classification/discriminant rule.
1) Generalization on the smallest decomposable components
2) Attribute removal
3) Concept tree Ascension
4) Vote propagation
5) Threshold control on each attribute
6) Threshold control on generalized relations
7) Rule transformation
8) Handling overlapping tuples

Relational database as resources for data mining for mining rules with attribute oriented induction can be read with data manipulation language select sql statement [20-22,25]. Using query for building rules presents efficient mechanism for understanding the mined rules [19,23]. Retrieve records from relational database with select sql statement is known but how to get and implement the simple select sql statement as to implement attribute oriented induction with simple select sql statement as easy and quick to get data result as the understanding.

Using threshold as a control for maximum number of tuples of the target class in the final generalized relation will not need anymore and as replacement group by operator in sql select statement will limit the final result generalization. Setting different threshold will generate different generalized tuples as the needed of global picture of induction repeatedly as time-consuming and tedious work [24]. All interesting generalized tuples as multiple rule can be generated as the global picture of induction by using group by operator or distinct function in sql select statement.

Build the logical formula as the representation final result for attribute oriented induction can not be done with select sql statement and not select sql statement capability to build the logical formula. But the sql statement can be matched with other applications like Java, Visual Basic, programming server program like ASP, JSP or PHP. The data results from sql statement can be used to create logical formula with that application software.

2. DATABASE DESIGN

As the connectivity with current or last research data example will refer to data in Cai [2] and Han et al. [8] as a concept hierarchy in figure 3 and example of data student in table 1. Figure 4 is the logical data model for database implementation where there are 5 tables, where table student as representative data from table1 dan other tables like Hierarchy_major, Hierarchy_Cat, Hierarchy_GPA and Hierarchy_Birth as the implementation from concept hierarchy in figure 1. Each concept tree from concept hierarchy will be transformed become a table. Database design in figure 4 is similar like star schema in Data Warehouse where table student as fact table and other tables as dimensional table. As a result multi dimensional concept in Data Warehouse can be applied where data can be roll up and drill down and data can be viewed in multiple dimensions with concept slice, dice and pivot [4,6,18]. Using aggregate count function and Group by operator in sql select statement will represent the roll up process [1,7].

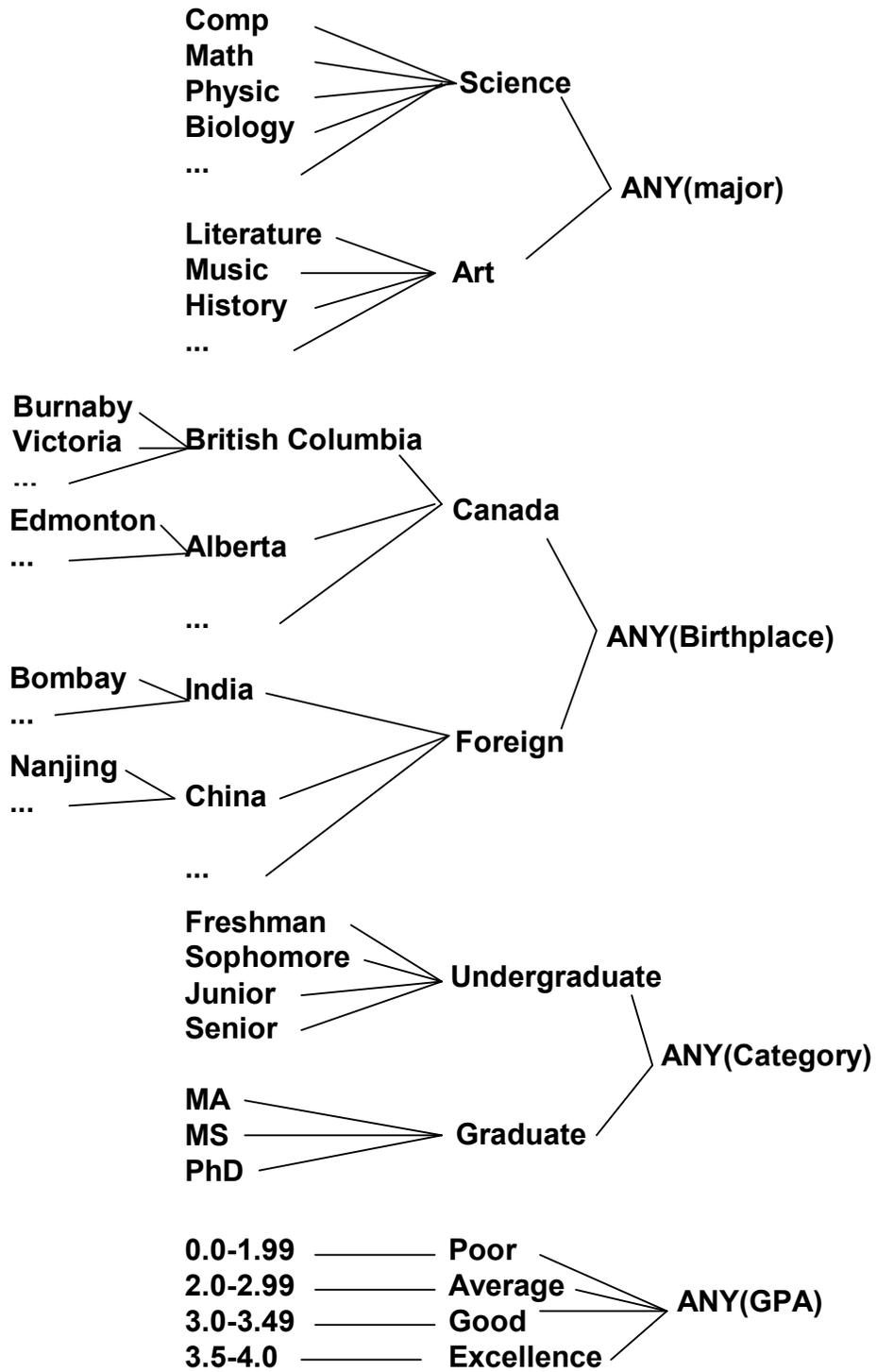

Fig. 3. A concept hierarchy table

Table I. Table Student

| Name | Category | Major | Birthplace | GPA |
|---|---|---|---|---|
| Anton | M.A. | History | Vancouver | 3.5 |
| Andi | Junior | Math | Calgary | 3.7 |
| Amin | Junior | Liberal arts | Edmonton | 2.6 |
| Anil | M.S. | Physics | Ottawa | 3.9 |
| Ayin | Ph.D. | Math | Bombay | 3.3 |
| Amir | Sophomore | Chemistry | Richmond | 2.7 |
| Acai | Senior | Computing | Victoria | 3.5 |
| Abdi | Ph.D. | Biology | Shanghai | 3.4 |
| Afun | Sophomore | Music | Burnaby | 3.0 |
| Agung | Ph.D. | Computing | Victoria | 3.8 |
| Ahing | M.S. | Statistics | Nanjing | 3.2 |
| Akuan | Freshman | literature | Toronto | 3.9 |

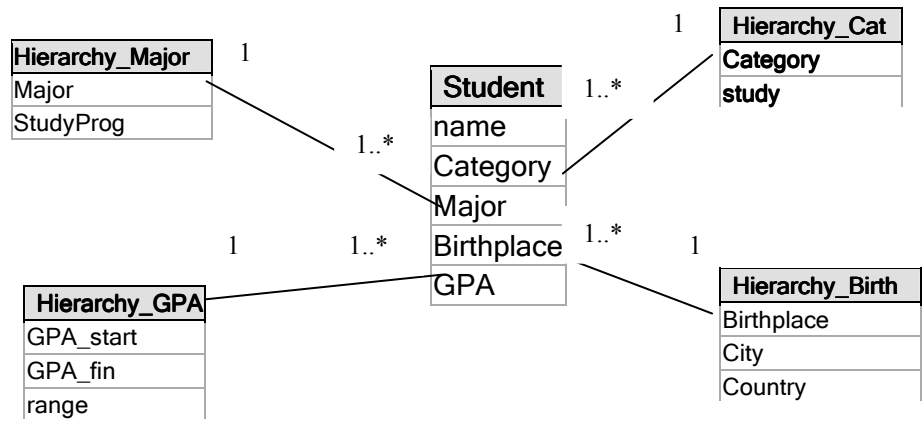

Fig. 4. Logical data model

3. CHARACTERISTIC RULE FOR GRADUATE STUDENT

For implementation the attribute oriented induction by doing the strategies steps, the the sql statement will be created based on example in in Cai [2] and Han et al. [8]. First we will find out characteristic rule for graduate student. The next sql statement will create table 2 as the selection for graduate student.

> select a.* from student a, hierarchy_cat b
> where a.category=b.category and b.study='graduate';

Table II. Table Graduate Student

| Name | Category | Major | Birthplace | GPA |
|---|---|---|---|---|
| Anton | M.A. | History | Vancouver | 3.5 |
| Anil | M.S. | Physics | Ottawa | 3.9 |
| Ayin | Ph.D. | Math | Bombay | 3.3 |

| Abdi | Ph.D. | Biology | Shanghai | 3.4 |
| Agung | Ph.D. | Computing | Victoria | 3.8 |
| Ahing | M.S. | Statistics | Nanjing | 3.2 |

The next sql statement will explain for the next generalization strategy as the attribute removal, where field Name will be removed because there is no generalization for this field and field category as generalization for graduate characteristic. Table 3 as the result for the sql statement select a.major, a.birthplace,a.GPA from student a, hierarchy_cat b where a.category=b.category and b.study='graduate';

Table III. Attribute removal for graduate student characteristic rule

| Major | Birthplace | GPA |
|---|---|---|
| History | Vancouver | 3.5 |
| Physics | Ottawa | 3.9 |
| Math | Bombay | 3.3 |
| Biology | Shanghai | 3.4 |
| Computing | Victoria | 3.8 |
| Statistics | Nanjing | 3.2 |

After Attribute removal generalization strategy then the next generalization strategy step is concept tree ascension, where attribute will be substituted by high level generalization as the first or minimal generalization. Table 4 show the transformation where based on concept hierarchy in figure 3 which is implemented with tables in database. For example the record number 1 for major = history has high generalization to Art, the Birth place = Vancouver has high level generalization to British Columbia and GPA = 3.5 has high level generalization as excellent. All the generalization based on concept hierarchy in figure 3.

Table IV. First generalization graduate student characteristic rule

| Major | Birthplace | GPA | Major | Birthplace | GPA |
|---|---|---|---|---|---|
| History | Vancouver | 3.5 | Art | British Columbia | Excellent |
| Physics | Ottawa | 3.9 | Science | Ontario | Excellent |
| Math | Bombay | 3.3 | Science | India | Good |
| Biology | Shanghai | 3.4 | Science | China | Good |
| Computing | Victoria | 3.8 | Science | British Columbia | Excellent |
| Statistics | Nanjing | 3.2 | Science | China | Good |

There are overlapping for record 4 and 6 in the right side table 4 and two of these overlapping records will be combined become one record. Table 5 show the combination for overlapping records and table 5 can be created with sql statement:

select c.studyprog, d.city,e.range
from student a, hierarchy_cat b, hierarchy_major c, hierarchy_birth d, hierarchy_gpa e
where a.category=b.category and b.study='graduate' and

```
        a.major=c.major and a.birthplace=d.birthplace and
        a.gpa>=e.gpa_start and a.gpa<=e.gpa_fin
group by c.studyprog, d.city,e.range;
```

or with another sql statement :

```
select distinct(c.studyprog), d.city,e.range
from student a, hierarchy_cat b, hierarchy_major c, hierarchy_birth d, hierarchy_gpa e
where a.category=b.category and b.study='graduate' and
        a.major=c.major and a.birthplace=d.birthplace and
        a.gpa>=e.gpa_start and a.gpa<=e.gpa_fin
```

Table V. Concept tree ascension for graduate student characteristic rule

| Major   | Birthplace       | GPA       |
|---------|------------------|-----------|
| Art     | British Columbia | Excellent |
| Science | Ontario          | Excellent |
| Science | British columbia | Excellent |
| Science | India            | Good      |
| Science | China            | Good      |

The next step generalization strategy step is vote propagation where with records will be accumulated. Table 6 is a result as the same with table 5 but only adding with column vote. The next sql statement as will be used to produce table 6.

```
select c.studyprog, d.city,e.range, count(c.studyprog) as Vote
from student a, hierarchy_cat b, hierarchy_major c, hierarchy_birth d, hierarchy_gpa e
where a.category=b.category and b.study='graduate' and
        a.major=c.major and a.birthplace=d.birthplace and
        a.gpa>=e.gpa_start and a.gpa<=e.gpa_fin
group by c.studyprog, d.city,e.range;
```

Table VI. Vote propagation for graduate student characteristic rule

| Major   | Birthplace       | GPA       | Vote |
|---------|------------------|-----------|------|
| Art     | British Columbia | Excellent | 1    |
| Science | Ontario          | Excellent | 1    |
| Science | British columbia | Excellent | 1    |
| Science | India            | Good      | 1    |
| Science | China            | Good      | 2    |

The column Birthplace still can be generalized as in concept hierarchy in figure 3, table 7 show the generalization to the highest level. There are 2 overlapping record, first record 2 and 3 next record 4 and 5 in the right side table 7 and these overlapping records will be combined become two record as the result in table 8. The next sql statement is used to produce table 8.

```
select c.studyprog, d.country,e.range, count(studyprog) as Vote
from student a, hierarchy_cat b, hierarchy_major c,  hierarchy_birth d, hierarchy_gpa e
where a.category=b.category and b.study='graduate' and
        a.major=c.major and a.birthplace=d.birthplace and
        a.gpa>=e.gpa_start and a.gpa<=e.gpa_fin
group by c.studyprog, d.country,e.range;
```

Table VII. Next generalization for graduate student characteristic rule

| Major | Birthplace | GPA | Major | Birthplace | GPA |
|---|---|---|---|---|---|
| Art | British Columbia | Excellent | Art | Canada | Excellent |
| Science | Ontario | Excellent | Science | Canada | Excellent |
| Science | British columbia | Excellent | Science | Canada | Excellent |
| Science | India | Good | Science | Foreign | Good |
| Science | China | Good | Science | Foreign | Good |

Table VIII. Further generalization for graduate student characteristic rule

| Major | Birthplace | GPA | Vote |
|---|---|---|---|
| Art | Canada | Excellent | 1 |
| Science | Canada | Excellent | 2 |
| Science | Foreign | Good | 3 |

Simplification will be performed in table 8 by unioning the first two records as show in table 9. As concept hierarchy in figure 3 then {art, science} can be generalized to ANY as show in table 10.

Table IX. Simplification for graduate student characteristic rule

| Major | Birthplace | GPA | Vote |
|---|---|---|---|
| {art, science} | Canada | Excellent | 3 |
| Science | Foreign | Good | 3 |

Table X. Generalization simplification for graduate student characteristic rule

| Major | Birthplace | GPA | Vote | t-weight |
|---|---|---|---|---|
| ANY | Canada | Excellent | 3 | 50% |
| Science | Foreign | Good | 3 | 50% |

The sql statement stop until produce table 8 for table 9 and 10 will possible can be produced with application programming as to produce the logical formula in rule (1) as the last generalization strategy step. One of last column in table 10 is t-weight as measure the typicality of each record in the characteristic rule where 50%=3/(3+3)*100.

(1) V(x)=graduate(x)→ (Birthplace(x) Є Canada ∧ GPA(x) Є excellent) V
(Major(x) Є science ∧ Birthplace(x) Є Foreign ∧ GPA(x) Є good)

Rule (1) is as qualitative rule and can be quantitative rule with adding quantitative measurement.

## 4. CHARACTERISTIC RULE FOR UNDERGRADUATE STUDENT

As complement characteristic rule for graduate student then characteristic rule for undergraduate student will be find out. The next sql statement will produce table 11 as selection for undergraduate student.

select a.* from student a, hierarchy_cat b
where a.category=b.category and b.study='undergraduate'

Table XI. Undergraduate student

| Name | Category | Major | Birthplace | GPA |
|------|----------|-------|------------|-----|
| Andi | Junior | Math | Calgary | 3.7 |
| Amin | Junior | Liberal arts | Edmonton | 2.6 |
| Amir | Sophomore | Chemistry | Richmond | 2.7 |
| Acai | Senior | Computing | Victoria | 3.5 |
| Afun | Sophomore | Music | Burnaby | 3.0 |
| Akuan | Freshman | literature | Toronto | 3.9 |

The next sql statement will explain for the next generalization strategy as the attribute removal, where field Name will be removed because there is no generalization for this field and field category as generalization for undergraduate characteristic. Table 12 as the result for the sql statement

select a.major, a.birthplace,a.GPA from student a, hierarchy_cat b
where a.category=b.category and b.study='undergraduate';

Table XII. Attribute removal for undergraduate student characteristic rule

| Major | Birthplace | GPA |
|-------|------------|-----|
| Math | Calgary | 3.7 |
| Liberal arts | Edmonton | 2.6 |
| Chemistry | Richmond | 2.7 |
| Computing | Victoria | 3.5 |
| Music | Burnaby | 3.0 |
| Literature | Toronto | 3.9 |

After Attribute removal generalization strategy then the next generalization strategy step is concept tree ascension, where attribute will be substituted by high level generalization as the first or minimal generalization. Table 13 show the transformation where based on concept hierarchy in figure 3 which is implemented with tables in database. For example the record number 1 for major = Math has high generalization to Science, the Birth place = Calgary has high level generalization to Alberta and GPA = 3.7 has high level generalization as excellent. All the generalization based on concept hierarchy in figure 3.

Table XIII. First generalization undergraduate student characteristic rule

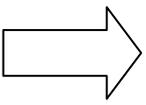

| Major | Birthplace | GPA |
|---|---|---|
| Math | Calgary | 3.7 |
| Liberal arts | Edmonton | 2.6 |
| Chemistry | Richmond | 2.7 |
| Computing | Victoria | 3.5 |
| Music | Burnaby | 3.0 |
| Literature | Toronto | 3.9 |

| Major | Birthplace | GPA |
|---|---|---|
| Science | Alberta | Excellent |
| Art | Alberta | Average |
| Science | British Columbia | Average |
| Science | British Columbia | Excellent |
| Art | British Columbia | Good |
| Art | Ontario | Excellent |

The next sql statement will produce result as in the right side table 13 where there is none overlapping records.

select c.studyprog, d.city,e.range
from student a, hierarchy_cat b, hierarchy_major c, hierarchy_birth d, hierarchy_gpa e
where a.category=b.category and b.study='undergraduate' and
     a.major=c.major and a.birthplace=d.birthplace and
     a.gpa>=e.gpa_start and a.gpa<=e.gpa_fin
group by c.studyprog, d.city,e.range;

or with another sql statement :

select distinct(c.studyprog), d.city,e.range
from student a, hierarchy_cat b, hierarchy_major c, hierarchy_birth d, hierarchy_gpa e
where a.category=b.category and b.study='undergraduate' and
     a.major=c.major and a.birthplace=d.birthplace and
     a.gpa>=e.gpa_start and a.gpa<=e.gpa_fin

The column Birthplace still can be generalized as in concept hierarchy in figure 3, table 14 show the generalization to the highest level. There are overlapping record for record 1 and 4 the overlapping records will be combined become one record as the result in table 15. The next sql statement is used to produce table 15.

select c.studyprog, d.country,e.range, count(studyprog) as Vote
from student a, hierarchy_cat b, hierarchy_major c, hierarchy_birth d, hierarchy_gpa e
where a.category=b.category and b.study='undergraduate' and
     a.major=c.major and a.birthplace=d.birthplace and
     a.gpa>=e.gpa_start and a.gpa<=e.gpa_fin
group by c.studyprog, d.country,e.range;

Table XIV. Next generalization for undergraduate student characteristic rule

| Major | Birthplace | GPA |
|---|---|---|
| Science | Alberta | Excellent |
| Art | Alberta | Average |
| Science | British Columbia | Average |
| Science | British Columbia | Excellent |
| Art | British Columbia | Good |
| Art | Ontario | Excellent |

| Major | Birthplace | GPA |
|---|---|---|
| Science | Canada | Excellent |
| Art | Canada | Average |
| Science | Canada | Average |
| Science | Canada | Excellent |
| Art | Canada | Good |
| Art | Canada | Excellent |

Table XV. Further generalization for undergraduate student characteristic rule

| Major | Birthplace | GPA | Vote |
|---|---|---|---|
| Science | Canada | Excellent | 2 |
| Art | Canada | Average | 1 |
| Science | Canada | Average | 1 |
| Art | Canada | Good | 1 |
| Art | Canada | Excellent | 1 |

Simplification will be performed in table 15 by unioning all records as show in table 16. As concept hierarchy in figure 3 then {art, science} can be generalized to ANY as show in table 17.

Table XVI. Simplification for graduate student characteristic rule

| Major | Birthplace | GPA | Vote |
|---|---|---|---|
| {Science,Art} | Canada | {Average, Good, Excellent} | 6 |

Table XVII. Generalization simplification for graduate student characteristic rule

| Major | Birthplace | GPA | Vote | t-weight |
|---|---|---|---|---|
| ANY | Canada | {Average, Good, Excellent} | 6 | 100% |

The sql statement stop until produce table 15 for table 16 and 17 will possible can be produced with application programming as to produce the logical formula in rule (2) as the last generalization strategy step. One of last column in table 17 is t-weight as measure the typicality of each record in the characteristic rule where 100%.

(2) V(x)=undergraduate(x)→ (Birthplace(x) Є Canada ∧
                GPA(x) Є {Average,Good, Excellent})
             [100%]

Rule (2) is as quantitative rule and can be qualitative rule by dropping quantitative measurement.

5. CLASSIFICATION/DISCRIMINATION RULE

The same with previous explanation for characteristic rule then classification/discriminant rule will be done based on data example in Cai [2] and Han et al. [8]. The next sql statement will produce table 18 based on generalization strategy steps.

```
select b.study,c.studyprog,d.city,e.range,count(*) as Vote
from student a, hierarchy_cat b, hierarchy_major c, hierarchy_birth d,
hierarchy_gpa e
where a.category=b.category and
      a.major=c.major and a.birthplace=d.birthplace and
      a.gpa>=e.gpa_start and a.gpa<=e.gpa_fin
group by b.study,c.studyprog,d.city,e.range
```

Table XVIII. Classification rule for student category

| study | studyprog | city | Range | Vote |
|---|---|---|---|---|
| graduate | Art | British Columbia | Excellent | 1 |
| graduate | Science | British Columbia | Excellent | 1 |
| graduate | Science | China | Good | 2 |
| graduate | Science | India | Good | 1 |
| graduate | Science | Ontario | Excellent | 1 |
| undergraduate | Art | Alberta | Average | 1 |
| undergraduate | Art | British Columbia | Good | 1 |
| undergraduate | Art | Ontario | Excellent | 1 |
| undergraduate | Science | Alberta | Excellent | 1 |
| undergraduate | Science | British Columbia | Average | 1 |
| undergraduate | Science | British Columbia | Excellent | 1 |

Record 1 until 5 is positive learning/target class and the last records 6 until 11 as negative learning/contrasting class. There are overlapping for record 2 and 11 and as generalization strategy step 8 then the records must be eliminated. The next sql statement will produce table 19.

```
select b.study,c.studyprog,d.country,e.range, count(*) as Vote
from student a, hierarchy_cat b, hierarchy_major c, hierarchy_birth d,
hierarchy_gpa e
where a.category=b.category and
      a.major=c.major and a.birthplace=d.birthplace and
      a.gpa>=e.gpa_start and a.gpa<=e.gpa_fin
group by b.study,c.studyprog,d.country,e.range
```

Table XIX. Final result classification rule for student category

| Study | studyprog | City | range | Vote |
|---|---|---|---|---|
| Graduate | Art | Canada | Excellent | 1 |
| Graduate | Science | Canada | Excellent | 2 |
| Graduate | Science | Foreign | Good | 3 |
| undergraduate | Art | Canada | Average | 1 |
| undergraduate | Art | Canada | Good | 1 |
| undergraduate | Art | Canada | Excellent | 1 |
| undergraduate | Science | Canada | Excellent | 2 |
| undergraduate | Science | Canada | Average | 1 |

Table 20 is final result with adding t-weight measurement the typicality of each tuple in the characteristic rule and d-weight as measurement the discriminating behavior of the learned classification rule which can be created with application software.

Table XX. Final result classification rule for student category

| Study | studyprog | City | range | Vote | t-weight | d-weight |
|---|---|---|---|---|---|---|
| Graduate | Art | Canada | Excellent | 1 | 1/(1+2+3)=16.67% | 1/(1+1)=50% |
| Graduate | Science | Canada | Excellent | 2 | 2/(1+2+3)=33.33% | 2/(2+2)=50% |
| Graduate | Science | Foreign | Good | 3 | 3/(1+2+3)=50% | 3/3=100% |
| undergraduate | Art | Canada | Average | 1 | 1/(1+1+1+2+1)=16.67% | 1/1=100% |
| undergraduate | Art | Canada | Good | 1 | 1/(1+1+1+2+1)=16.67% | 1/1=100% |
| undergraduate | Art | Canada | Excellent | 1 | 1/(1+1+1+2+1)=16.67% | 1/(1+1)=50% |
| undergraduate | Science | Canada | Excellent | 2 | 2/(1+1+1+2+1)=33.33% | 2/(2+2)=50% |
| undergraduate | Science | Canada | Average | 1 | 1/(1+1+1+2+1)=16.67% | 1/1=100% |

Rule (3) is logical formula for graduate student and rule (4) as logical formula for undergraduate student which can be created with application software.

(3) V(x)=graduate(x)→
    (Major(x) Є Art Λ Birthplace(x) Є Canada Λ GPA(x) Є Excellent) [50%]  V
    (Major(x) Є science Λ Birthplace(x) Є Canada Λ GPA(x) Є Excellent) [50%]  V
    (Major(x) Є science Λ Birthplace(x) Є Foreign Λ GPA(x) Є Good) [100%]

(4) V(x)=undergraduate(x)→
    (Major(x) Є Art Λ Birthplace(x) Є Canada Λ GPA(x) Є Average) [100%] V
    (Major(x) Є Art Λ Birthplace(x) Є Canada Λ GPA(x) Є Good) [100%]  V
    (Major(x) Є Art Λ Birthplace(x) Є Canada Λ GPA(x) Є Excellent) [50%]  V
    (Major(x) Є science Λ Birthplace(x) Є Canada Λ GPA(x) Є Excellent) [50%]  V

(Major(x) Є science ∧ Birthplace(x) Є Foreign ∧ GPA(x) Є Average) [100%]

6. CONCLUSION

By using sql statement for producing table 19 we can produce characteristic rule for graduate student, characteristic rule for undergraduate student and classification/discriminant rule for student simultaneously. Particularly for characteristic rule for further generalization in order to make the last characteristic rule result then performance application software will be needed, specifically for make t-weight, d-weight and logical formula as rules.

Sql statement is the shortest, easy and simple way to get characteristic and classification/discriminant rule from relational database. The powerful sql statement will be increased with application software helping by doing any others sql statement can not be done.

Database design for concept hierarchy as a part of attribute oriented induction will influence the process for sql statement. The knowledge for transformation concept hierarchy will be needed as a basic foundation to do the best select sql statement implementation by transform each of concept tree in concept hierarchy become a table for searching characteristic or classification/discriminant rule from relational database as data mining process.